\documentclass[aps,pra,floatfix,twocolumn,showpacs]{revtex4}
\usepackage{dcolumn}% Align table columns on decimal point
\usepackage{bm}
\usepackage{soul}
\usepackage{color}
\usepackage{bm}
\usepackage{graphicx} \usepackage{amsmath} \usepackage{amssymb}
\newcommand{\comment}[1]{}
\newcommand{\lp}{\underline{p}}
\newcommand{\lo}{\underline{\omega}}
\newcommand{\up}{\overline{p}}
\newcommand{\uo}{\overline{\omega}}
\renewcommand{\b}{\beta}

\newcommand{\be}{\begin{eqnarray}}
\newcommand{\ee}{\end{eqnarray}}

\newcommand{\no}{\label}

\begin{document}

\title{Excluding joint probabilities from quantum theory}

\author{Armen E. Allahverdyan and Arshag Danageozian}
\affiliation{Yerevan Physics Institute, 2 Alikhanian Brothers street, Yerevan
0036, Armenia}

\begin{abstract}
  
Quantum theory does not provide a unique definition for the joint
probability of two non-commuting observables, which is the next
important question after the Born's probability for a single observable.
Instead, various definitions were suggested, e.g. via
quasi-probabilities or via hidden-variable theories. After reviewing
open issues of the joint probability, we relate it to quantum
imprecise probabilities, which are non-contextual and are consistent
with all constraints expected from a quantum probability.  We study two
non-commuting observables in a two-dimensional Hilbert space and show
that there is no precise joint probability that applies for any quantum
state and is consistent with imprecise probabilities. This contrasts to
theorems by Bell and Kochen-Specker that exclude joint probabilities for
more than two non-commuting observables, in Hilbert space with dimension
larger than two.  If measurement contexts are included into the
definition, joint probabilities are not anymore excluded, but they are
still constrained by imprecise probabilities. 

\end{abstract}


\maketitle

Open problems of quantum mechanics revolve around the notions of
non-commutativity and probability
\cite{deMuynck,Busch,abn,leifer,fine_prl,fine_jmp,fine_brit,bugaj}. In
contrast to classical mechanics, where introducing probability relates
to a limited control over the experimental situation, the quantum
probability presents itself as a fundamental description of a single
quantum observable, as well as pairs of commuting observables
\cite{deMuynck,Busch,abn}. However, the structure and interpretation of the
joint probability for non-commuting observables is an open problem,
primarily because the standard machinery of quantum mechanics precludes
a direct and precise joint measurement of such observables
\cite{deMuynck,Busch,abn}. This need not exclude the definition of joint
probability, since the latter may turn out to be a construct recovered
indirectly through different measurements, or may even point out to a
generalized theory beyond quantum mechanics. As seen below, many
possible candidates for the joint probability were proposed. This
subject was active since the inception of quantum mechanics
\cite{hillery}, and it is much alive now with the needs of quantum
information and foundations \cite{ferrie}. Theorems by Bell and
Kochen-Specker demonstrate the non-existence of the joint probability
for more than two variables living in a Hilbert space ${\cal H}$ with dimension
${\rm dim}{\cal H}>2$ \cite{fine_prl,fine_jmp,accardi,gudder_bell}. 
The Kochen-Specker theorem looks for a set of observables
for which no joint probability exists for any state, while Bell's
theorem is restricted to specific (entangled) states. Neither of these
theorems restricts joint probability for two non-commuting observables,
which is the next important case after the probability of a single
observable. Recently, Malley added to the Kochen-Specker set-up the
consistency with quantum conditional probability, which is derived
assuming (additionally) the projection postulate \cite{malley}. This
modified set-up disallows joint probabilities for any pair of
non-commuting observables for ${\rm dim}{\cal H}>2$ \cite{malley}.

Here we show in which sense the joint probability can be excluded
(without assuming the projection postulate) already for the minimal
situation, {\it viz.} any pair of non-commuting observables living in a
two-dimensional Hilbert space. Note that the projection postulate is not
necessary for the empiric applicability of quantum probability; e.g.
there are interpretations of quantum mechanics that do not employ it
\cite{evo,grif}. We also study how our results depend on the
standard assumption of a single probability space for non-commuting
observables. 

% we extend the refutation for more general (e.g. Dempster-Shafer) joint
% probability models. 

% Still it is a serious issue that the direct empirical support
% for the joint probability is so far lacking. Hence there is a
% conceptually paradoxical situation, because the physics does need a
% joint statistical description of non-commuting observables, as
% evidenced by a huge effort since the inception of quantum mechanics.

We illustrate the existing approaches towards defining joint
probabilities for two non-commuting observables (Hermitian operators)
$A$ and $B$ ($[A,B]\not =0$) living in a Hilbert space ${\cal H}$.  Let
$P_a$ ($Q_b$) be the eigen-projector of $A$ ($B$) that corresponds to
the eigenvalue $a$ ($b$). 

Our first condition is that a hypothetical joint probability $p_{ab}$ of
eigenvalues of $a$ and $b$ in a state with density matrix $\rho$ holds
marginality conditions:
\begin{eqnarray}
  \label{eq:9}
{\sum}_a\, p_{ab}={\rm tr}(\rho Q_b), \qquad 
{\sum}_b\, p_{ab}={\rm tr}(\rho P_a), 
\end{eqnarray}
where the first (second) summation is taken over all {\it different}
eigenvalues of $A$ ($B$). 

We embedded the joint probabilities of non-commuting observables into a
single space; cf.~(\ref{eq:9}). This assumption can be questioned, since
non-commuting $P_a$ and $Q_b$ demand different measurement contexts
\cite{szabo}.  But we adopt it (till the last section), since it is a
common point for all approaches|including the Bell and Kochen-Specker
theorems|that look for the joint probability \cite{fine_prl,fine_jmp}.
Below we give examples for a joint probability. 

{\it (i)} The most known approach is that of quasi-probabilities
\cite{hillery,ferrie}. They are linear over $\rho$, but they
ought to become negative for certain states \cite{hillery,ferrie}. This
limits their interpretation as probabilities \cite{mueck}. In addition,
there are many different quasi-probabilities and it is not clear which
one of them applies in a concrete situation, even if it is positive.
Despite of these issues, quasi-probabilities are widely used; e.g. in
quasi-classics \cite{hillery}, information theory \cite{ferrie}, signal
processing \cite{cohen,cohen_zap}, and statistical mechanics \cite{work,mat}.
A good example of quasi-probability is the Terletsky-Margenau-Hill
function \cite{ter}:
\begin{eqnarray}
  \label{eq:17}
p_{ab}^{\rm \bf T}= {\rm tr}(\rho\, (P_aQ_b+Q_bP_a)\,)/2.
\end{eqnarray}
$p_{ab}^{\rm\bf T}$ is negative for certain
$\rho$'s, since $P_aQ_b+Q_bP_a$ has a negative eigenvalue if
$[P_a,Q_b]\not=0$. Eq.~(\ref{eq:17}) relates to weak-values \cite{bamboo} and its
negativity has a physical meaning \cite{mat}. 
$p_{ab}^{\rm \bf T}$ is more convenient than the Wigner's function that
is defined only for specific pairs of observables. 

The negativity of (\ref{eq:17}) for certain $\rho$'s has deeper roots:
any probability defined via  
\begin{eqnarray}
  \label{eq:177}
p_{ab}^{\rm \bf C}= {\rm tr}(\rho\, \Pi_{ab})\geq 0, ~~{\rm due~to}~~ \Pi_{ab}\geq  0,
\end{eqnarray}
which holds (\ref{eq:9}) for any $\rho$ (hence $\sum_b\Pi_{ab}=P_a$
and $\sum_a\Pi_{ab}=Q_b$) and which is forced to be positive by the
non-negative definiteness of Hermitian $\Pi_{ab}$, will imply that $A$ and $B$
are commuting: $AB=BA$ \cite{fine_jmp,fine_brit}.

{\it (ii)} The linearity of (\ref{eq:17}, \ref{eq:177}) over $\rho$ has
two implications. First, it means that within the standard measurement
theory $p_{ab}^{\rm \bf T}$ can be determined via measuring Hermitian
$P_aQ_b+Q_bP_a$ in a state with an unknown $\rho$ \cite{deMuynck,abn}.
Second, consider the process of mixing: out of two ensembles with
different $\rho_1$ and $\rho_2$, one makes up a new ensemble by taking
$\rho_k$ with probability $\mu_k$ ($k=1,2$). The new ensemble has
density matrix $\rho_{\rm mix}=\sum_{k=1}^2\mu_k\rho_k$.  All quantities
that are linear over density matrix (e.g. $p_{ab}^{\rm \bf T}$) will
depend directly on $\rho_{\rm mix}$ keeping no memory on separate
preparations $\rho_k$ that contributed into the mixing. 

But the linearity is not obligatory: one can
define a joint probability via sacrificing the linearity over $\rho$,
so as to ensure the positivity for all $\rho$, keeping correct
marginals as in (\ref{eq:9}) \cite{cohen_zap}. Such constructs cannot
be measured via the standard approach, but they are determined
theoretically once $\rho$ is known. They are not unique,
but they were employed for studying complementarity \cite{cohen}. The
simplest {\it non-linear} example holding (\ref{eq:9}) is
\begin{eqnarray}
  \label{eq:10}
p_{ab}^{\rm \bf N}={\rm tr}( P_a\sqrt{\rho}\,Q_b\sqrt{\rho}\,)\geq 0.
\end{eqnarray}

{\it (iii)} Deterministic hidden-variable theories offer another
definition of the joint probability \cite{mermin}.  Here is an example
based on a hidden-variable theory proposed by Bell for ${\rm dim}\,{\cal
H}=2$ \cite{mermin}. Recall the Bloch representation for any
non-negative operator $R\geq 0$ (density matrix or projector) with ${\rm
tr}\,R=1$ in ${\rm dim}\,{\cal H}=2$:
\begin{eqnarray}
  \label{eq:11}
R=(1+\vec{\b}_R\,\vec{\sigma})/2,\quad \vec{\b}_R={\rm tr}(R\vec{\sigma}), 
\quad |\vec{\b}_R|\leq 1,
\end{eqnarray}
where $\vec{\b}_R=(\b_{R,x},\b_{R,y},\b_{R,z})$ is a real vector and
$\vec{\sigma}$ is the vector of Pauli matrices. Now the hidden-variable is
a real vector $\vec{m}$ with $|\vec{m}|=1$. For any projector $P$ we
recover Born's rule via the integration over $\vec{m}$ with the
characteristic function
$\vartheta[\,\vec{\b}_P(\vec{\b}_\rho+\vec{m})]$ \cite{mermin}:
\begin{eqnarray}
  \label{eq:12}
  {\rm tr}(\rho P)=\frac{1}{2}(1+\vec{\b}_\rho
  \vec{\b}_P)=\int\frac{{\rm
      d}\vec{m}}{4\pi}\,\vartheta[\,\vec{\b}_P(\vec{\b}_\rho+\vec{m})], 
\end{eqnarray}
where $\vartheta[x]$ is the step-function ($\vartheta[x\geq 0]=1$ and
$\vartheta[x<0]=0$), and $\int{\rm d}\vec{m}$ integrates over all
directions of 3-dimensional hidden-variable space. Eq.~(\ref{eq:12}) is
verified by going to spherical coordinates: ${\rm
  d}\vec{m}=\sin(\theta)\,{\rm d} \theta\, {\rm d}\phi$. This
model suggests the joint probability which holds (\ref{eq:9}):
\begin{eqnarray}
  \label{eq:120}
p_{ab}^{\rm \bf B}=  \int\frac{{\rm
      d}\vec{m}}{4\pi}\,\vartheta[\,\vec{\b}_{P_a}(\vec{\b}_\rho+\vec{m})]\,
  \vartheta[\,\vec{\b}_{Q_b}(\vec{\b}_\rho+\vec{m})]\geq 0.
\end{eqnarray}

Now (\ref{eq:17}) and (\ref{eq:10}) are non-contextual definitions, i.e.
$p_{ab}$ depends only on $P_a$ and $Q_b$, and not on other projectors
$\sum_{a}P_a=I$ and $\sum_{a}Q_b=I$ of $A$ and $B$ ($I$ is the unity
operator on ${\cal H}$).  Eq.~(\ref{eq:120}) is also non-contextual, but
its generalizations to ${\rm dim}\,{\cal H}>2$ ought to be contextual
\cite{gudder}. 

\comment{ One can object against (\ref{eq:12}, \ref{eq:120}), e.g.
because it does not hold the single-particle Bell inequalities derived
for $\rho\propto I$ \cite{gudder_bell}. Below we exclude (\ref{eq:120})
from weaker premises. }

Besides (\ref{eq:9}) there is another natural condition to
which a physical joint probability $p_{ab}$ should satisfy 
\cite{gudder_bell,accardi}:
\be
\label{tot}
&&p_{ab}={\rm tr}(\rho P_aQ_b)~~{\rm if}~\\
&&[P_a, \rho]=0 ~~{\rm or}~~ [\rho, Q_b]=0 ~~ {\rm or}~~
[P_a, Q_b]=0. 
  \label{eq:5} 
\ee 
To explain (\ref{tot}, \ref{eq:5}), note that (\ref{eq:5}) forces
${\rm tr}(\rho P_aQ_b)$ to have features of joint probability, i.e. it is
symmetric with respect to $P_a$ and $Q_b$, 
non-negative and holds (\ref{eq:9}). Imposing (\ref{tot}, \ref{eq:5})
is especially obvious for $[P_a, Q_b]=0$, where $P_aQ_b$ is a projector.
For $[P_a, \rho]=0$, ${\rm tr}(\rho P_aQ_b)$ can be recovered
as the average of an Hermitian observable $(P_aQ_b+Q_bP_a)/2$. Alternatively,
we note that measuring $P_a$ does not change $\rho$
statistically. Hence the joint probability is found by first measuring
$P$ and then $Q$: ${\rm tr}(\rho P_a Q_b)={\rm tr}(P_a\rho P_a Q_b)$.
Likewise for $[Q_b,\rho]=0$. 

Now (\ref{eq:10}) and (\ref{eq:120})|which are non-negative for any
$\rho$|do not hold (\ref{tot}, \ref{eq:5}). Eq.~(\ref{eq:10}) does not
hold the third condition in (\ref{eq:5}), while (\ref{eq:120}) does not
hold the first and second conditions in (\ref{eq:5}), as seen already in
the simplest case $\rho=1/2$. Eq.~(\ref{eq:17}) holds (\ref{eq:9},
\ref{tot}, \ref{eq:5}), but it is negative for certain $\rho$'s.  
In this context we formulate the following 

{\it Conjecture}: there is {\it no} joint probability
$p_{ab}(P_a,Q_b,\rho)$ that is non-negative for any $\rho$ (i.e.
quasi-probabilities are excluded), is non-contextual and holds
(\ref{tot}, \ref{eq:5}); cf.~\cite{bugaj}. We stress that we do not
require $p_{ab}(P_a,Q_b,\rho)$ to be linear over $\rho$. This conjecture
is yet to be (in)validated. 

Below we show that joint probabilities can be excluded from a different
argument that relates to imprecise probabilities.  In contrast to the
usual joint probability, the imprecise probabilities are well-defined
given conditions of non-contextuality and correspondence with the
commutative situation. The physical reason for this is that there exists
a specific type of quantum uncertainty for two non-commuting observables
that is captured by the quantum imprecise probability, which is
consistent with all conditions expected from a quantum probability. 

Before continuing, we comment on joint measurements schemes for
non-commuting observables \cite{deMuynck,Busch,arthurs,abn_prl}, a known
method for characterizing non-commutativity. Generally, this method does
not provide definitions for joint probabilities that are new compared
with the above analysis. In particular, it is uclear to which extent the
existing schemes for joint measurements produce intrinsic results that
characterize the system itself, and not approximate measurements
employed \cite{uffink}; e.g. they do not hold condition (\ref{eq:9}) of
the joint probability \cite{uffink}. Instead, they focus on different
conditions, e.g. the unbiasedness \cite{arthurs} or stability
\cite{abn_prl}. 

{\bf Projectors} are self-adjoint operators $P$ with $P^2=P$. Any projector
$P$ in a Hilbert space ${\cal H}$ bijectively relates to the sub-space
${\cal S}_P$ of ${\cal H}$ \cite{jauch_book}: 
\be
\no{bu}
{\cal S}_P=\{ |\psi\rangle \in {\cal H};~ P|\psi\rangle=|\psi\rangle \}.
\ee
Eigenvalues of $P$ are $0$ and/or $1$, and it is a quantum analogue of the
characteristic function for a classical set \cite{jauch_book}. Hence
projectors define quantum probability: with a density
matrix $\rho$, the probability of finding the eigenvalue $1$
of $P$ is given by Born's formula ${\rm tr}(\rho P)$.

The simplest projectors are $0$ and $I$. We define
\be
P\geq P'~~{\rm means}~~\langle \psi|P-P'|\psi\rangle\geq 0~~{\rm for~any}~~|\psi\rangle.
\no{grum}
\ee
Now apply (\ref{grum}) with $|\psi\rangle=|\psi_0\rangle$, where $P|\psi_0\rangle=0$,
and then with $|\psi\rangle=|\psi_1\rangle$, where $P'|\psi_1\rangle=|\psi_1\rangle$.
Hence the eigenvalues of $P$ and $P'$ relate to each other leading to
\be
\no{sar}
PP'=P'P=P'~~{\rm if}~~ P\geq P'.
\ee
Projectors (generally non-commuting) support logical operations
\cite{jauch_book}. {\it Negation} $P^\perp=I-P$, 
is a projector that has zeros (ones) whenever $P$ has ones (zeroes). 
{\it Conjunction} $P\land Q$ contains only those
vectors that belong both to ${\cal S}_{P}$ and ${\cal S}_{Q}$. 
Thus ${\cal S}_{P\land Q}={\cal S}_{Q}\cap {\cal S}_{P}$.
{\it Disjunction} $P\lor Q$ cannot be defined via ${\cal S}_{Q}\cup {\cal S}_{P}$ 
(set-theoretic union), because the latter is not a Hilbert (linear) space.
The minimal Hilbert space that contains ${\cal S}_{Q}\cup {\cal S}_{P}$,
is made of all linear combinations of the vectors from ${\cal S}_{Q}\cup {\cal S}_{P}$: 
\begin{gather}
\label{uruk}
{\cal S}_{P\lor Q}=\{ |\psi\rangle_P+|\psi\rangle_Q;\, |\psi\rangle_P \in {\cal S}_P, |\psi\rangle_Q \in {\cal S}_Q \}.
\end{gather}
There are alternative representations \cite{jauch_book}:
\begin{align}
  \label{bars}
  P\land Q={\rm max}_R[\, R\,|\,R^2=R, \, R\leq P, \, R\leq Q\,], \\
  P\lor Q={\rm  min}_R[\, R\,|\,R^2=R, \, R\geq P, \, R\geq Q\,],
  \label{colibri}
\end{align}
where the maximization and minimization go over projectors $R$
\cite{jauch_book}. Indeed, if $R\leq P$, and $R\leq Q$ in (\ref{bars}),
then due to (\ref{bu},\ref{sar}), ${\cal S}_R$ is a subspace of both
${\cal S}_P$ and ${\cal S}_Q$. The maximal such subspace is ${\cal
S}_{P\land Q}$. Likewise, if $R\geq P$, and $R\geq Q$, then ${\cal S}_R$
has to contain both ${\cal S}_P$ and ${\cal S}_Q$. The minimal such
subspace is ${\cal S}_{P\lor Q}$ as (\ref{uruk}) shows.

Now $P\lor Q=0$ only if $P=Q=0$, but $P\land Q$ can be zero also for
non-zero $P$ and $Q$; e.g. for non-zero $P$ and $Q$ in ${\rm dim}{\cal
H}=2$, we have either $P=Q$ or $P\land Q=0$ (and then $P\lor Q=I$). 

The above three operations are related with each other and with a limiting 
process \cite{jauch_book}:
\be
P\lor Q= (P^\perp\land Q^\perp)^\perp, ~~
P\land Q={\rm lim}_{n\to\infty}(PQ)^n.
\label{fox}
\ee
They are well-known in quantum logics \cite{jauch_book}, but we shall
employ them without a specific logical interpretation. Eqs.~(\ref{bars},
\ref{colibri}) were generalized to non-negative operators
\cite{moreland}. For $[P,Q]\equiv PQ-QP=0$ we have from
(\ref{sar}--\ref{fox}) ordinary features of classical characteristic
functions:
\be
P\land Q=PQ,\quad P\lor Q=P+Q-PQ.
\no{dum}
\ee
 
\comment{Commuting projectors are simultaneously measurable.  With a density
matrix $\rho$, the probability of finding the eigenvalue $1$ (resp. $0$)
of $P$ is given by Born's formula ${\rm tr}(\rho P)$ (resp.  ${\rm
tr}(\rho P^\perp)$). }

{\bf Imprecise classical probability} generalizes the usual (precise)
probabilities \cite{intro}: the measure of uncertainty for an
event $E$ is an interval $[\lp(E),\up(E)]$, where
$0\leq\lp(E)\leq \up(E)$ are called lower and upper probabilities,
respectively. Now $\lp(E)$ (resp. $1-\up(E)$) is a measure of a sure
evidence in favor (resp. against) of $E$. The event $E$ is surely more
probable than $E'$, if $\lp(E)\geq \up(E')$. The usual probability is
recovered for $\lp(E)=\up(E)$. Two different pairs $[\lp(E),\up(E)]$
and $[\lp'(E),\up'(E)]$ can hold simultaneously (i.e they are {\it
  consistent}) if
\begin{eqnarray}
  \label{eq:100}
\lp'(E)\leq \lp(E)~~{\rm and}~~
\up'(E)\geq \up(E).  
\end{eqnarray}
Every $[\lp(E),\up(E)]$ is consistent with $\lp'(E)=0$, $\up'(E)=1$.
This non-informative situation is not described by the usual
theory that inadequately offers for it 
the homogeneous probability \cite{intro}. Now $[\lp(E),\up(E)]$ does
not imply that there is an explicit (but possibly unknown) precise
probability for $E$ located in between $\lp(E)$ and $\up(E)$
\cite{papa}. 

There are various imprecise classical probability theories
\cite{intro}, from a rather weak structures called upper and lower
measures in \cite{suppes_zanotti_erken} to the Dempster-Shafer imprecise
probability \cite{dempster,shafer}. The latter has numerous
applications e.g. in decision making and artificial intelligence
\cite{intro}. Recently it was applied for describing aspects of the
Bell's inequality \cite{suppes_zanotti,vourdas,fletcher}.  The quantum
imprecise probability is to be sought independently, along the physical
arguments. Below we recall how it is determined. 

{\bf Imprecise joint quantum probability} is sought for two non-commuting
projectors $P$ and $Q$. We look for upper $\uo(P,Q)$ and lower
$\lo(P,Q)$ non-negative probability operators. The respective upper and
lower probabilities in a state with density matrix $\rho$ are given by
Born's rule:
\begin{eqnarray}
  \label{borno}
\up(P,Q)={\rm
  tr}(\rho\,\uo(P,Q)), \quad \lp(P,Q)={\rm tr}(\rho\,\lo(P,Q)).  
\end{eqnarray}
The linearity of $\up(P,Q)$ and $\lp(P,Q)$ over $\rho$ can be motivated
as in {\it (ii)} above. We determine $\uo(P,Q)$ and $\lo(P,Q)$ from
conditions (\ref{eq:1}--\ref{eq:4}) \cite{armen}
\begin{align}
  \label{eq:1}
&  0\leq \lo(P, Q)=\lo(Q, P)\leq \uo(P, Q)=\uo(Q, P)\leq
I. \\ \nonumber\\
  \label{eq:2}
&  [\,\omega(P, Q),Q\,]=[\,\omega(P, Q),P\,]=0 ~{\rm for}~ \omega=\lo,\uo.\\ 
\nonumber\\  
  \label{eq:3}
&  \lo(P, Q)= \uo(P, Q)=P Q~~~ {\rm if}~~~ [P, Q]=0. \\ \nonumber\\
  \label{eq:4} & {\rm tr}(\rho \,\lo(P, Q)\,)\leq {\rm tr}(\rho\, P
Q)\leq{\rm tr}(\rho\,\uo(P, Q)\,)~ {\rm if}~ (\ref{eq:5})~{\rm holds}.
\end{align} 
Eqs.~(\ref{eq:1}, \ref{borno}) force $0\leq\lp(P,Q)\leq
\up(P,Q)\leq 1$ for any $\rho$. Eq.~(\ref{eq:1}) also
demands symmetry with respect to $P$ and $Q$ that is natural for the 
joint probability.  

Now $\lo(P, Q)$ and $\uo(P, Q)$ are non-contextual in the sense that
they depend only on $P$ and $Q$. Even a stronger feature holds:
(\ref{eq:2}) shows that both $\lo(P, Q)$ and $\uo(P, Q)$ can be measured
together with either $P$ or $Q$; see also (\ref{eq:18}). Hence the
imprecise probability of two non-commuting observables does not lead to
additional non-commutativity \cite{f2}. 

\comment{Otherwise for a joint probability of two observables we shall need
another two non-commuting observables, and so on leading to an infinite
regress. }

For $[P,Q]=0$ we revert to the usual joint probability; see
(\ref{eq:3}). For $Q=I$ we get from (\ref{eq:3}, \ref{borno}) the
marginal and precise probability of $P$.  Eqs.~(\ref{eq:4}, \ref{eq:5})
also refer to the consistency with the precise probability
[cf.~(\ref{eq:100})], because the latter is well-defined not only for
$[P,Q]=0$, but also under conditions (\ref{eq:5}), where it amounts to
${\rm tr}(\rho P Q)$. 

Eqs.~(\ref{eq:1}--\ref{eq:4}) suffice for deducing \cite{armen}:
\begin{eqnarray}
  \label{eq:20}
  \lo(P, Q)=P\land Q, \qquad   \uo(P, Q)=P\lor Q-(P-Q)^2,~
\end{eqnarray}
where $\lo(P, Q)$ and $\uo(P, Q)$ are (resp.) the largest and the smallest positive
operators holding (\ref{eq:1}--\ref{eq:4}). Now $\lo(P, Q)$ is a
projector, while $\uo(P, Q)$ is generally just a non-negative
operator. For $[P,Q]=0$, we have from (\ref{dum}, \ref{eq:20}):
$\lo(P, Q)=\uo(P, Q)=PQ$, as required by (\ref{eq:3}). 

Eqs.~(\ref{eq:20}) imply (\ref{eq:2}), because|as follows from (\ref{bars}, 
\ref{colibri}) and checked directly|$P\land Q$, $P\lor Q$ and
$(P-Q)^2$ commute with each other and with $P$ and $Q$. Hence
\begin{eqnarray}
  \label{eq:18}
  [\,\uo(P, Q),\,\lo(P, Q)\,]=0. 
\end{eqnarray}
The origin of (\ref{eq:20}) is understood from (\ref{eq:1}, \ref{eq:2})
and (\ref{bars}, \ref{colibri}), i.e. $P\land Q$ and $P\lor Q$ qualify
as certain (resp.) lower and upper probability operators, while the
factor $(P-Q)^2$ in (\ref{eq:20}) is needed to ensure (\ref{eq:3}). 

In (\ref{borno}) we stress that if we would search for
imprecise probability without assuming the linear dependence on $\rho$
(but still assuming the analogues of (\ref{eq:1}--\ref{eq:4})), we can
obtain only more precise (in the sense of (\ref{eq:100})) quantities
than $\up(P,Q)$ and $\lp(P,Q)$ in (\ref{borno}).  The same argument is
given for (\ref{eq:2}): relaxing it (but keeping other
features) will lead to more precise probability.  Thus conclusions obtained
via linear $\up(P,Q)$ and $\lp(P,Q)$ will stay intact. 

{\bf A geometric feature} of (\ref{eq:20}) is that both
$PQP$ (i.e. the restriction of $Q$ into ${\cal S}_P$) and 
$QPQ$ hold:
\begin{eqnarray}
  \label{eq:22}
  \lo(P,Q)\leq PQP,\, QPQ\,\leq   \uo(P,Q).
\end{eqnarray}
Now $\lo(P,Q)\leq PQP$ is shown from $P\land Q\leq Q$ [see (\ref{bars})]
that implies $P\land Q=P(P\land Q)P\leq PQP$.  And $PQP\leq\uo(P,Q)$
follows from: $\uo(P,Q)-PQP=\uo(P,Q) -P\uo(P,Q)P=\uo(P,Q)(I-P)\geq 0$
recalling that $[\uo(P,Q),P]=0$ from (\ref{eq:2}, \ref{eq:20}).  

Eqs.~(\ref{eq:1}, \ref{eq:4}) can be deduced from (\ref{eq:20},
\ref{eq:22}), which also imply a version of sub- and
super-additivity:
\begin{eqnarray}
  \label{chin}
{\sum}_a\uo(P_a,Q)\geq Q\geq {\sum}_a\lo(P_a,Q),\quad {\sum}_a P_a=I.
\end{eqnarray} 
Thus the additive marginalization leads to an
upper bound $\sum_a\uo(P_a,Q)$ (and lower bound $\sum_a\lo(P_a,Q)$) for
the correct marginal probability $\uo(I,Q)=\lo(I,Q)$. We also note from
(\ref{chin}) that non-negative operators $\uo(P_a,Q_b)$ and
$\lo(P_a,Q_b)$ do not hold a semi-spectra resolution, e.g.
$\sum_{a,b}\uo(P_a,Q_b)\geq I$, hence they cannot be interpreted via
generalized measurements \cite{deMuynck,Busch}. Note as well that the monotonicity does not
hold: $\uo(P,Q)\not\leq\uo(I,Q)=Q$, though $P\leq I$;
cf.~(\ref{eq:13}). 

{\bf Consistency with quantum conditional probabilities}.  Two-time
(conditional) quantum probabilities are defined as follows
\cite{deMuynck}: first measure $P$ and assume the validity of the projection
postulate.  Now the result $P=1$ implies the post-measurement density matrix
$P\rho P/{\rm tr}(\rho P)$. Then measuring $Q$ leads to conditional
probability $p_{Q=1|P=1}={\rm tr}(\rho PQP)/{\rm tr}(\rho P)$ for the
$Q=1$. Likewise, we obtain $p_{P=1|Q=1}={\rm tr}(\rho QPQ)/{\rm tr}(\rho
Q)$ when measuring first $Q$ and then $P$. The two-time probabilities
are not usual conditional probabilities, since they do not lead to joint
probabilities, e.g.  applying the usual formulas does not generally lead
to unique results due to
$p_{P=1|Q=1}\,\,{\rm tr}(\rho Q)={\rm tr}(\rho QPQ) \not = {\rm tr}(\rho PQP)$.
However, imprecise joint probabilities (\ref{borno}) can lead to
conditional imprecise probabilities. They are defined via the usual
formulas, because the marginal (i.e. ${\rm tr}(\rho P)$ and ${\rm
tr}(\rho Q)$) probabilities are precise. 
Eqs.~(\ref{eq:22}) show that two-time probabilities are bound by the
conditional imprecise probability, e.g. 
\begin{align}
\label{doka}
\frac{{\rm tr}(\rho \lo(P,Q))}{{\rm tr}(\rho P)}
\leq \frac{{\rm tr}(\rho PQP)}{{\rm tr}(\rho P)}\leq
\frac{{\rm tr}(\rho \uo(P,Q))}{{\rm tr}(\rho P)}.
\end{align}

\comment{
Eqs.~(\ref{eq:20}) imply the basic requirements of the imprecise
probability ($K$ is a projector):
\begin{align}
  \label{eq:7}
& \lo(P, Q+K)\geq \lo(P, Q)+\lo(P, K) ~~{\rm if}~~ QK=0, \\
  \label{eq:8}
& \uo(P, Q+K)\leq \uo(P, Q)+\uo(P, K) ~~{\rm if}~~ QK=0.
\end{align}
Eqs.~(\ref{eq:7}, \ref{eq:8}) generalize the additivity of the precise
probability for incompatible events (which means $QK=0$ for
projectors). The additivity is still meaningful, but it leads to a
lower bound for the (true) lower probability and an upper bound for
the upper probability; see (\ref{eq:7}, \ref{eq:8}), respectively.}

{\bf Inconsistency with precise joint probabilities}. We saw above that
imprecise probabilities (\ref{borno}) are consistent (in the sense of
(\ref{eq:100})) with all instances, where quantum mechanics provides
reasonable definitions of joint or conditional probability;
cf.~(\ref{eq:4}, \ref{doka}). Hence we {\it assume} that the reasonable
definition of precise quantum joint probability should also be
consistent with (\ref{borno}). 

Thus we study a joint probability $p_{ab}$ under two conditions. First
we require (\ref{eq:9}). Second, we demand that $p_{ab}$ is consistent
with (\ref{eq:20}) in the sense of (\ref{eq:100}) for any density matrix
$\rho$ and all $a$ and $b$:
\begin{eqnarray}
  \label{eq:70}
0\leq  {\rm tr}(\rho \,\lo(P_a, Q_b)\,)
\leq p_{ab}\leq {\rm tr}(\rho \,\uo(P_a, Q_b)\,).
\end{eqnarray}
We do not demand that $p_{ab}$ depends only on $P_a$, $Q_b$; i.e. for
$p_{ab}$ we allow contextuality. We also do not demand that its
dependence on $\rho$ is linear. Since for $p_{ab}$ we require
$p_{ab}\geq 0$ for all $\rho$, quasi-probabilities are naturally
excluded from consideration.  Any theory that generalizes quantum
mechanics and predicts joint probability for any preparation will be
constrained by (\ref{eq:70}). Note that (\ref{eq:70}) is not stronger or
weaker than (\ref{tot}, \ref{eq:5}). 

Let two non-commuting observables $A$ and $B$ with (resp.)
eigen-projectors $P_1+P_2=I$ and $Q_1+Q_2=I$ live in a two-dimensional
Hilbert space. Due to $P_a\lor Q_b=I$ and $P_a\land Q_b=0$,
(\ref{eq:20}) simplify as:
\begin{eqnarray}
  \label{eq:13}
  \lo(P_a,Q_b)=0, ~~~~
\uo(P_a,Q_b)={\rm tr}(P_aQ_b),
\end{eqnarray}
i.e. both probability operators reduce to numbers \cite{f1}.

Our main result is that for given non-commutative $A$ and $B$,
there is a density matrix that violates (\ref{eq:70}) or (\ref{eq:9}).
In this sense the precise joint probability does not exist already in
${\rm dim}{\cal H}=2$. Indeed, take $p_{22}\leq {\rm tr}(P_2Q_2)={\rm
tr}(P_1Q_1)$ from (\ref{eq:70}, \ref{eq:13}), $p_{21}\leq {\rm
tr}(\rho Q_1)$ from (\ref{eq:9}), and employ them in $p_{22}+p_{21}=
1-{\rm tr}(\rho P_1)$ from (\ref{eq:9}):
\begin{eqnarray}
  \label{eq:14}
{\rm tr}(P_1Q_1) +{\rm tr}(\rho P_1) +{\rm tr}(\rho
Q_1)-1\geq 0.
\end{eqnarray}
For given $P_1$ and $Q_1$, there is a density matrix $\rho$ for which
the left-hand-side of (\ref{eq:14}) is negative. Indeed, its
positivity amounts in the Bloch representation (\ref{eq:11}) to
$\vec{\b}_{P_1}\vec{\b}_{Q_1}
+\vec{\b}_{\rho}(\vec{\b}_{P_1}+\vec{\b}_{Q_1})\geq -1$, and it can be
violated e.g. as follows. If $\vec{\b}_{P_1}\vec{\b}_{Q_1}=\cos\alpha$
then we can choose $\vec{\b}_{\rho}\to 1$ and
$\vec{\b}_{\rho}\vec{\b}_{P_1}=\vec{\b}_{\rho}\vec{\b}_{Q_1}
=-\cos\frac{\alpha}{2} $
producing $\cos\alpha-2\cos\frac{\alpha}{2}<-1$ for $0<\alpha<\pi$,
i.e. for those values of $\alpha$, where $[A,B]\not=0$. 

\comment{Interestingly, (\ref{eq:14}) has the same mathematical
structure as single-particle Bell inequalities found in
\cite{gudder,accardi}. }

Thus, no precise joint probability is consistent with the
quantum imprecise probability for all states.

{\bf Summary and outlook.} We defined a set-up of searching for a joint
probability for two non-commuting observables. This is the next problem
after the Born's probability for a single observable. An open aspect of
this problem was formulated as a conjecture. We show that there is
no joint probability for two non-commuting observables in a
two-dimensional Hilbert space, if should this probability apply to any
state and should be consistent with the quantum imprecise probability. 
This statement does not apply, if
measurement contexts are introduced. Now we
look for the joint probability generalizing out 
condition (\ref{eq:9}). Redefine the Born's
probabilities as conditional ones $p_{a|P_a}={\rm tr}(\rho P_a)$ and
$p_{b|Q_b}={\rm tr}(\rho Q_b)$, where conditioning can account for
different devices needed to measure $P_a$ and $Q_b$. We cannot deduce
$p_{a|P_a}$ and $p_{b|Q_b}$ from a single joint probability, i.e. marginality condition
(\ref{eq:9}) does not apply anymore, and is generalized as follows.  
We postulate two different joint
probabilities $p_{ab|P_a}$ and $p_{ab|Q_b}$ holding
${\sum}_a p_{ab|Q_b}={\rm tr}(\rho Q_b)$ and 
${\sum}_b p_{ab|P_a}={\rm tr}(\rho P_a)$,
where ${\sum}_b p_{ab|Q_b}=p_{a|Q_b}$ and
${\sum}_a p_{ab|P_a}=p_{b|P_a}$ are well-defined, but they need not be given
by Born's formulas. Eqs.~(\ref{eq:2}, \ref{eq:18}) mean that
$\lo(P_a,Q_b)$ and $\uo(P_a,Q_b)$ can be measured simultaneously with
each other and with either $P_a$ or $Q_b$. Hence they can be used in both
contexts $P_a$ and $Q_b$. We constrain $p_{ab|P_a}$ and
$p_{ab|Q_b}$ demanding their consistency with (\ref{borno})
for any $\rho$ [cf.~(\ref{eq:13}, \ref{eq:70})]:
${\rm tr}(\rho\, \lo(P_aQ_b)) \leq  p_{ab|P_a},\,\,
p_{ab|Q_b}\leq {\rm tr}(\rho\, \uo(P_aQ_b))$.
Together with generalized marginality these conditions define a
set of probabilities that contains $\hat{p}_{ab|P_a}= {\rm tr}(\rho P_aQ_bP_a)$ and
$\hat{p}_{ab|Q_b}= {\rm tr}(\rho Q_bP_aQ_b)$; cf.~(\ref{eq:22}). 
Elsewhere we shall explore this approach, noting that 
descriptions via sets of probabilities are well-known in mathematical statistics
\cite{intro}. 

\comment{\begin{equation}
  \label{eq:160}
 {\rm tr}(\rho\, \lo(P_aQ_b)) \leq  p_{ab|P_a},\,\,
p_{ab|Q_b}\leq {\rm tr}(\rho\, \uo(P_aQ_b)).
\end{equation}
}


\begin{thebibliography}{99}

\bibitem{deMuynck} W.M. de Muynck, {\it Foundations of quantum
    mechanics, an empiricist approach} (Kluwer Academic Publishers,
  Dordrecht, 2002).

\bibitem{Busch} P. Busch, P. J. Lahti and P. Mittelstaedt, {\it The
    Quantum Theory of Measurement} (Springer, Berlin, 1996).

\bibitem{abn} A. E. Allahverdyan, R. Balian and Th. M. Nieuwenhuizen,
  Phys. Rep. {\bf 525}, 1 (2013).  

  % -166 Understanding quantum measurement from the solution of
  % dynamical models

\bibitem{leifer} D. Jennings and M. Leifer, Contemporary Physics, {\bf
    57}, 60 (2016).

%-82 No Return to Classical Reality


\bibitem{fine_prl} 
A. Fine, Phys. Rev. Lett. {\bf 48}, 291 (1982).

\bibitem{fine_jmp} 
A. Fine, J. Math. Phys. {\bf 23}, 1306 (1982).

% Joint distributions, quantum correlations, and commuting observables

\bibitem{fine_brit} 
A. Fine, Brit. J. Phil. Soc. {\bf 24}, 1 (1973).


\bibitem{bugaj}
S. Bugajski, Z. Naturforsch. A {\bf 33}, 1383 (1978).  
%-1385 A Remark on Joint Probability Distributions for Quantum Observables


\bibitem{hillery} M.Hillery, R.F. O'Connell, M.O. Scully and
  E.P. Wigner, Phys. Rep. {\bf 106}, 121 (1984).

% \bibitem{tatar}V.I.
% Tatarskii, Soviet Physics Uspekhi {\bf 26}, 311 (1983).

% The Wigner representation of quantum mechanics

\bibitem{ferrie} C. Ferrie, Rep. Prog. Phys. {\bf 74}, 116001 (2011).

  % Quasi-probability representations of quantum theory with
  % applications to quantum information science

\bibitem{accardi}  L. Accardi and A. Fedullo, 
Lett. Nuovo Cimento {\bf 34}, 161  (1982).

%-172 On the Statistical Meaning of the Complex Numbers in Quantum Mechanics

\bibitem{gudder_bell} S. P. Gudder, Found. Phys. {\bf 14}, 997 (1984).

\bibitem{malley} J. D. Malley, Phys. Rev. A {\bf 69}, 022118 (2004).

\bibitem{evo} F.J. Tipler, PNAS, {\bf 111}, 11281 (2014).

% Quantum nonlocality does not exist, 

\bibitem{grif} R.B. Griffiths, Found. Phys. {\bf 41}, 705 (2011).

\bibitem{szabo}  L.E. Szabo, Found. Phys. Lett. 
{\bf 8}, 417 (1995). 
% -436 Is quantum mechanics compatible with a deterministic universe? Two
% interpretations of quantum probabilities
G. Bana and T. Durt, {\it ibid.}  {\bf 27} 1355 (1997).
% -1373 Proof of Kolmogorovian censorship
Th.M. Nieuwenhuizen, {\it Where Bell went wrong}, arXiv:0812.3058.
D.A. Slavnov, Theoretical and mathematical physics, {\bf 136}, 1273
(2003).

%-1279 Quantum measurements and Kolmogorov's theory of probability

\bibitem{mueck}W. M\"uckenheim, Phys. Rep. {\bf 133}, 337 (1986).
  J. Maddox, Nature {\bf 320}, 481 (1986). D.E. Parry, {\it ibid.}
  {\bf 321}, 644 (1986). W. M\"uckenheim, {\it ibid.}  {\bf 324}, 307
  (1986).


\bibitem{cohen} L. Cohen, {\it Time Frequency Analysis: Theory and
    Applications} (Prentice-Hall, NJ, 1995).

\bibitem{cohen_zap}L. Cohen and Y. I. Zaparovanny
J. Math. Phys. {\bf 21}, 794 (1980). 

\bibitem{work} A.E. Allahverdyan, Phys. Rev. E {\bf 90}, 032137 (2014).

\bibitem{mat} M. Lostaglio, Phys. Rev. Lett. {\bf 120}, 040602 (2018).
%Quantum fluctuation theorems, contextuality and work quasi-probabilities


\bibitem{ter} J. G. Kirkwood, Physical Review {\bf 44}, 31
(1933). Y. P.  Terletsky, Journ. Exper. Theor. Phys. {\bf 7}, 1290 (1937).
%-1298 The limiting transition from quantum to classical mechanics
P.A.M. Dirac, Rev. Mod. Phys. {\bf 17}, 195 (1945). 
H. Margenau and R. N. Hill, Progress of Theoretical Physics {\bf
26}, 722 (1961). A. O. Barut, Phys. Rev. {\bf 108}, 565 (1957).

\bibitem{bamboo}J.B. Hartle, Phys. Rev. A {\bf 70}, 022104 (2004).
% Linear positivity and virtual probability
C. Bamber and J. S. Lundeen, Phys. Rev. Lett. {\bf 112}, 070405 (2014).
L.M. Johansen, arXiv:0804.4379. 
% {\it A unique quasi-probability for projective yes-no measurements}

%\bibitem{pusey} M. F. Pusey, Phys. Rev. Lett. {\bf 113}, 200401 (2014).


\bibitem{mermin}N.D. Mermin, Rev. Mod. Phys. {\bf 65}, 803 (1993).

\bibitem{gudder}
S. P. Gudder, J. Math. Phys. {\bf 11}, 431 (1970).
% On HiddenVariable Theories

\bibitem{arthurs} E. Arthurs and M.S. Goodman, Phys. Rev. Lett. 
  {\bf 60}, 2447 (1988).
  
\bibitem{abn_prl} A. E. Allahverdyan, R. Balian and Th.M. Nieuwenhuizen,
Phys. Rev. Lett. {\bf 92}, 120402, (2004).  

% Determining a quantum state by means of a single apparatus

\bibitem{uffink}J. Uffink, Int. J. Theor.  Phys. {\bf 33}, 199 (1994).

%\bibitem{bumbarash} {\it FAQs of Imprecise Probability}, available at
%https://msse.gatech.edu/research/interval/\\WhyImpreciseProbability.html

\bibitem{intro}{\it Introduction to Imprecise Probabilities} ed. by
  T. Augustin, F.P.A. Coolen, G. de Cooman and M.C.M. Troffaes (J.
  Wiley \& Sons, 2014).

%\bibitem{huber}P.J. Huber, {\it Robust Statistics} (Wiley \& Sons, NY, 1981). 

\bibitem{suppes_zanotti_erken}P. Suppes and M. Zanotti, Erkenntnis, {\bf
    31}, 323-345 (1989).

\bibitem{dempster} A.P. Dempster, Ann. Math. Stat. {\bf 38}, 325 (1967).

% -339 Upper and Lower Probabilities Induced by a Multivalued Mapping

\bibitem{shafer} G. Shafer, {\it The Mathematical Theory of Evidence}
(Princeton University Press, Princeton 1976). 

\bibitem{suppes_zanotti}P. Suppes and M. Zanotti, Found. Phys. {\bf
    21}, 1479 (1991).

\bibitem{vourdas}A. Vourdas, J. Math. Phys. {\bf 55}, 082107 (2014).

  % Quantum probabilities as Dempster-Shafer probabilities in the
  % lattice of subspaces

\bibitem{fletcher}
B.H. Feintzeig and S.C. Fletcher, Found. Phys. {\bf 47}, 294 (2017) 
%On Noncontextual, Non-Kolmogorovian Hidden Variable Theories

\bibitem{armen}  A.E. Allahverdyan, New J. Phys. {\bf 17}, 085005 (2015).

%Imprecise probability for non-commuting observables


%\bibitem{jauch} J. M. Jauch, Synthese {\bf 29}, 131 (1974). 

  % It was Jauch who used a theorem on the intersections of the ranges
  % of the spectral projections of momentum and coordinate operators to
  % point out the nonexistence of a well-behaved joint probability
  % distribution. This "Jauch theorem" (J) (which will be reported in
  % Sec. II) represents a much more intuitive and very strong
  % explication of the incompatibility of coordinate and momentum.

\bibitem{jauch_book} J.M. Jauch, {\it Foundations of Quantum
    Mechanics} (Addison-Wesley, Reading, 1968).


%\bibitem{qlog}{\it Handbook of quantum logics and quantum structures} ed. by 
%K. Engesser, D.M. Gabbay, and D. Lehmann (Elsevier, Amsterdam, 2007).

\bibitem{moreland}
T. Moreland and S. Gudder, Lin. Alg. Appl. {\bf 286}, 1 (1999).
H. Du, C. Deng, and Q. Li, Sci. China A {\bf 49}, 545 (2006).

% -17 Infima of Hilbert space effects, 

% -556 On the infimum problem of Hilbert space effects




% \bibitem{jauch_piron} J. M. Jauch and C. Piron, Helv. Phys. Acta {\bf
%     36}, 827 (1963).

\bibitem{papa}A. Papamarcou and T.L. Fine, The Annals of Probability,
  {\bf 14}, 710-723 (1986).

\bibitem{f2} Another possibility to require a joint measurability is to
insist on $\underline{\omega}(P ,Q)+\overline{\omega}(P ,Q)+P\leq 1$.
Then the averages over $\rho$ of $\underline{\omega}(P ,Q)$,
$\overline{\omega}(P ,Q)$, and $P$ can be found via a single POVM
measurement \cite{deMuynck}. This condition cannot work already for
$PQ=QP$, as seen from (\ref{eq:2}). 

\bibitem{f1} For $P\to Q$, $\uo(P,Q)$ jumps from $\approx I$ to $P$
  for $P=Q$; cf.~(\ref{eq:3}). A similar jumping from $0$ to $P$ is
  seen for $\lo(P,Q)$; see (\ref{eq:13}). Such jumps present no
  problems, since $[\lp=0,\up=1]$ is consistent with any other
  probability. 



\end{thebibliography}
\end{document}